\documentclass[journal=jpclcd,manuscript=letter]{achemso}

\usepackage{chemformula} % Formula subscripts using \ch{}
\usepackage[T1]{fontenc} % Use modern font encodings
\usepackage{graphicx}

\author{Carlos Marante}
\affiliation{Department of Physics, University of Central Florida, Orlando, Florida 32816, USA}
\author{Lina Frans\'{e}n}
\affiliation{Nantes Universit\'{e}, CNRS, CEISAM UMR 6230, F-44300 Nantes, France}
\author{Alexie Boyer}
\affiliation{Universite Claude Bernard Lyon 1, CNRS, Institut Lumière Matière, F-69622 Villeurbanne, France}
\author{Vincent Loriot}
\affiliation{Universite Claude Bernard Lyon 1, CNRS, Institut Lumière Matière, F-69622 Villeurbanne, France}
\author{Franck L\'{e}pine}
\affiliation{Universite Claude Bernard Lyon 1, CNRS, Institut Lumière Matière, F-69622 Villeurbanne, France}
\author{Luca Argenti}
\email{luca.argenti@ucf.edu}
\affiliation{Department of Physics, University of Central Florida, Orlando, Florida 32816, USA}
\author{Morgane Vacher}
\email{morgane.vacher@univ-nantes.fr}
\affiliation{Nantes Universit\'{e}, CNRS, CEISAM UMR 6230, F-44300 Nantes, France}
\author{Saikat Nandi}
\email{saikat.nandi@univ-lyon1.fr}
\affiliation{Universite Claude Bernard Lyon 1, CNRS, Institut Lumière Matière, F-69622 Villeurbanne, France}

\title{Femtosecond non-adiabatic confinement of molecular dication yield}

\keywords{multi-photon ionization, atto-chemistry, non-adiabatic dynamics}

\begin{document}

\begin{tocentry}

\includegraphics[width=\linewidth]{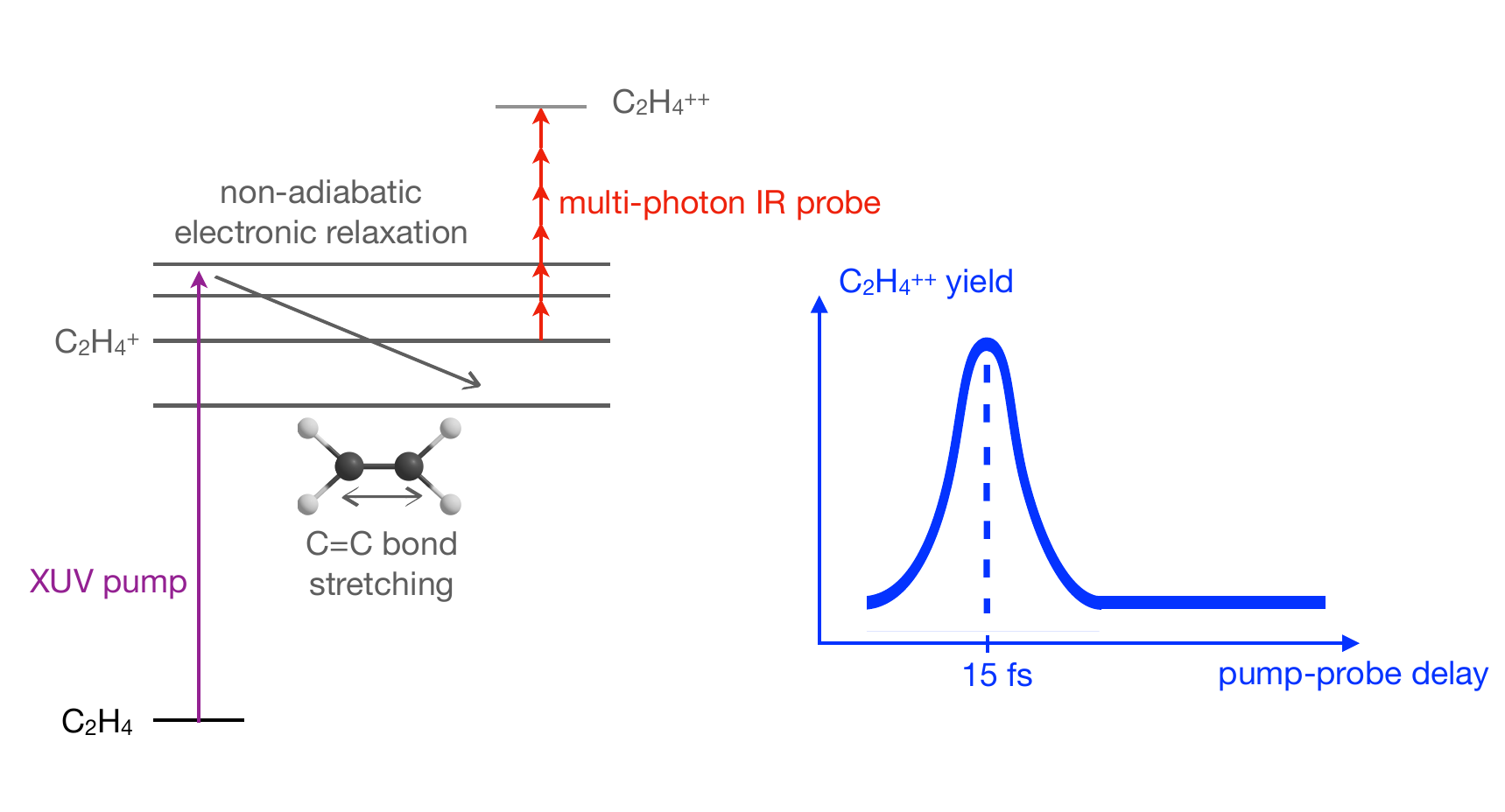}

\end{tocentry}

\begin{abstract}
Doubly charged molecular cations often carry signatures of electronic correlation and electron-nuclear entanglement present in the parent cation. Here, we produce ethylene dications using a combination of extreme ultraviolet pump and near-infrared probe pulses, observing a peak in the dication yield at a pump-probe delay of approximately 15 femtoseconds. Ab-initio calculations, which explicitly take into account coupled electron-nuclear dynamics induced by the pump and the multi-photon nature of the probe-induced ionization step, reproduced the observed delay in the yield. It originates from resonant enhancement of the multi-photon ionization of the electronically excited ethylene cation as the carbon--carbon double bond expands. However, this effect is tempered by rapid non-adiabatic relaxation of the excited ionic states. Our results suggest a general mechanism whereby ultrafast non-adiabatic relaxation of a molecular ion can compete with its strong-field ionization rate, confining the dication yield to a narrow temporal window of a few femtoseconds.
\end{abstract}

Doubly charged molecular ions usually dissociate into two singly charged fragments due to the repulsive Coulombic interaction between the two charges\cite{stapelfeldt1998}. When the dication survives long enough to be detected, they often provide a unique way to study the interplay between electronic and nuclear degrees of freedom in the parent molecular cation. Over the past few decades, synchrotron radiation has been extensively used to study processes that generate doubly charged species, such as photo-double ionization\cite{weber2004,bolognesi2004} and Auger-Meitner decay\cite{becker1996}. However, the stationary character of these measurements limits the information on the intermediate excited singly-charged ions. The availability of ultrashort (few hundreds of attoseconds to few tens of femtoseconds) pulses with photon energies in the extreme-ultraviolet (XUV) and the x-ray domain\cite{constant2025}, has opened up the possibility to detect dication and probe the molecular photoionization dynamics in a time-resolved manner. For example, non-adiabatic relaxation dynamics in cations of polycyclic aromatic hydrocarbons were studied by measuring the XUV pump -- near-infrared (NIR) probe delay-dependent yield of their respective dications\cite{marciniak2015}. These species were further investigated in relation to correlation bands in polyatomic systems, a prominent signature of electron-electron correlation\cite{herve2021}. Doubly charged ions from a nucleobase, adenine, have provided snapshots of charge migration in the parent cation\cite{mansson2021}. Ultrafast isomerization in acetylene dication was observed following ionization by femtosecond x-ray pulses\cite{liekhus2015}. Signatures of few-femtosecond electron transfer processes in XUV-ionized amino-benzene were extracted from the corresponding dication yield as a function of the XUV pump -- NIR probe delay\cite{vismarra2024}. Altogether, these studies highlight the role of the time-dependent molecular dication yield in revealing the ultrafast dynamics in the XUV-ionized parent cation. However, in these cases the influence of the multi-photon NIR probe is often overlooked, due to the challenges in simulating the probing step\cite{herve2021,vismarra2024}.  

\begin{figure*}[!htb]
\centering
\includegraphics[width=1.0\linewidth]{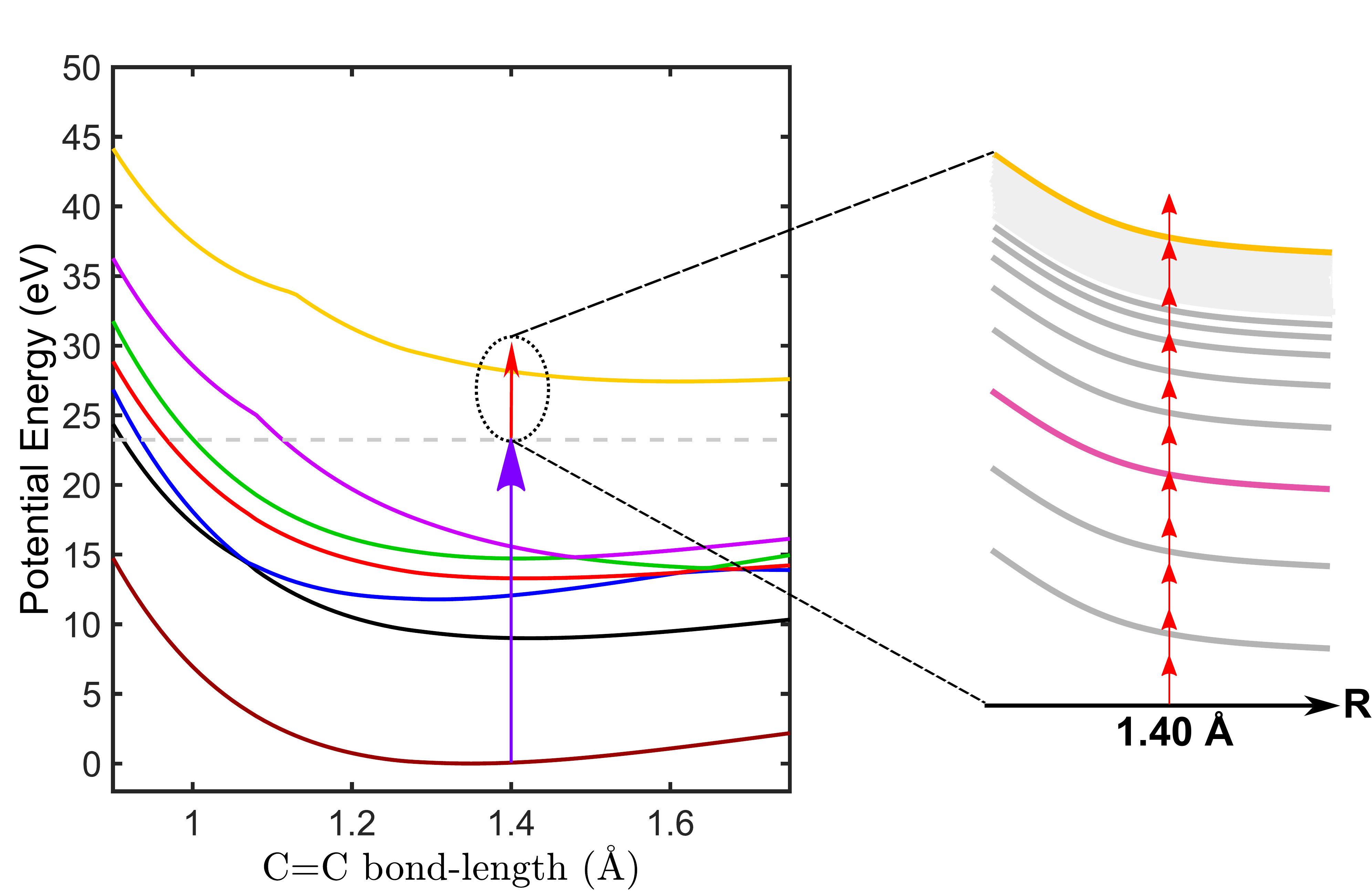}
\caption{The potential energy curves (solid lines) for various electronic states in ethylene as a function of the C=C bond length (left-panel): neutral ground state (maroon), D$_0$ (black), D$_1$ (blue), D$_2$ (red), D$_3$ (green), D$_4$ (purple) in $^{13}$C$^{12}$CH$_4^+$ cation and S$_0$ (yellow) in $^{13}$C$^{12}$CH$_4^{++}$ dication. These adiabatic potential energy curves have been computed at the SA-CASSCF level using the ANO-RCC-VDZP basis set, a (12,12) active space for the neutral species, (11,12) for the cation and (10,12) for the dication. The geometry was set to be D$_{2h}$ and only the C=C distance was varied. The vertical purple arrow indicates the XUV-pump pulse. The horizontal dashed line indicates the maximum photon energy reachable by the pump, which is below the photo-double ionization threshold for ethylene. To reach the ground state in the dication, we used a multi-photon NIR probe pulse (denoted by the red arrow). It requires around $10$ to $11$ NIR photons (see the right panel for a simplified pictorial depiction) to reach the S$_0$ state in the ethylene dication from the cation prepared by the XUV-pump. Due to its multi-photon nature, the NIR-probe can initiate a resonance-enhanced multi-photon transition (REMPI), via an electronic state (magenta) below the ground state of the dication, increasing the yield.}
\label{fig1}
\end{figure*}

Here, we investigate the formation of doubly charged ethylene ions using an XUV pump -- NIR probe scheme, where ultrafast non-adiabatic dynamics in the XUV-ionized ethylene cation compete with multi-photon ionization driven by the NIR (central wavelength $\sim 800$ nm,  photon energy $= 1.55$ eV) probe, directly influencing the resulting dication yield. An attosecond pulse train acting as the pump photoionizes and excites the molecule, while the absorption of several NIR photons from the time-delayed probe ejects a second electron from the singly-charged ion resulting in the dication (see Fig.~\ref{fig1} for a pictorial depiction). The photon energy of the XUV pulse is limited between $17$ and $23.3$ eV (see the dashed horizontal line in Fig.~\ref{fig1}), which is significantly lower than the photo-double ionization threshold of the molecule\cite{gaire2014}. Doubly ionized ethylene ions have a lifetime as long as few hundred nanoseconds\cite{jochim2017}, making them an ideal benchmark for examining fragmentation dynamics in polyatomic systems following strong-field ionization with few-femtosecond NIR pulses\cite{xie2012,wells2013,xie2014,larimian2016}. In the present case, we observe the ethylene dication yield to be strongly peaked around $15$~fs pump-probe delay. Starting from a neutral ground state, the broadband XUV pump induces non-adiabatic dynamics by simultaneously ionizing and exciting the molecule to several cationic states, from D$_0$ to D$_4$ (see Fig.~\ref{fig1}). Advanced trajectory surface hopping calculations\cite{vacher2022,fransen2024} indicate that the population of the D$_1$ state also peaks at around $15$~fs. To understand the effect of the intense NIR probe that ultimately leads to the production of the dication, we explicitly simulated the multi-photon ionization using a recent implementation of the transition-density-matrix time-dependent close-coupling formalism\cite{randazzo2023}. This approach of combining the non-adiabatic dynamics following the first ionization step with the non-perturbative ionization of the intermediate ion was essential to interpret the experimental findings. At a specific pump-probe delay, the length of carbon-carbon double bond in the singly charged cation undergoing non-adiabatic relaxation increases, allowing the possibility to reach the ground state of the dication via resonance enhanced multi-photon ionization (REMPI) processes. On one hand, the ionization yield of an electronic state can increase as a function of the pump-probe delay, while on the other hand, its electronic population decreases due to non-adiabatic relaxation, with D$_1$ and D$_2$ being ultimately responsible for the largest contribution to the peak of the dication yield at around $15$~fs. This competition between increased ionization rate as the molecular ionic structure relaxes and the onset of ultrafast non-adiabatic electronic relaxation dynamics is expected to play a role in other molecules, especially when using high-energy XUV or, x-ray photons as pump, thus simultaneously ionizing and exciting the system \cite{herve2021,mansson2021,vismarra2024,lee2021,barillot2021}.

\begin{figure*}[!htb]
\centering
\includegraphics[width=0.94\linewidth]{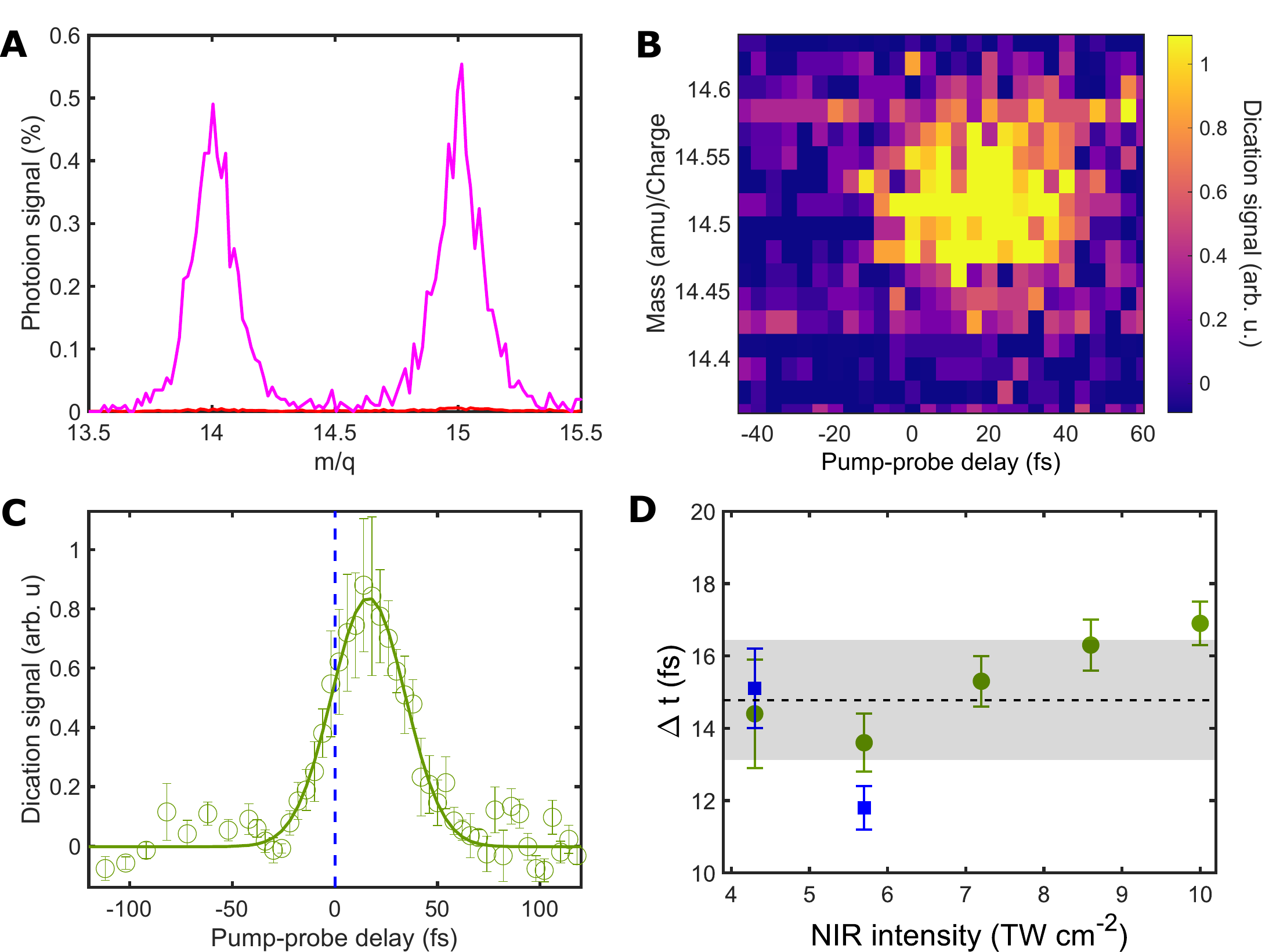}
\caption{({\bf A}) XUV-only (magenta, solid line) and NIR-only (red, solid line) time-of-flight mass spectra. In both cases, the signal has been normalized with respect to the corresponding parent cation ($^{13}$C$^{12}$CH$_4^+$) yield. The NIR-only mass spectra was collected at an approximate intensity of $8.6$ TW~cm$^{-2}$. No contributions from the dication can be observed in either of the spectra. ({\bf B}) Typical two-color signal for the dication, $^{13}$C$^{12}$CH$_4^{++}$. ({\bf C}) Same as in ({\bf B}), but integrated over the corresponding mass window. The error bars indicate statistical fluctuations. The solid line represents least-squares fitting with a time-delayed Gaussian function (see Supporting Information for details). ({\bf D}) Shift $\Delta t$ in temporal appearance of the dication relative to the pump-probe zero-delay. The uncertainties come from the fitting. The dashed line represents weighted average of the individual values, with the shaded area indicating the $99\%$ confidence interval.  Circles (squares): XUV pump pulses via HHG in xenon (krypton).}
\label{fig2}
\end{figure*}

Doubly-charged ethylene ions, $^{12}$C$_2$H$_4^{++}$, can dissociate symmetrically leading to the production of $^{12}$CH$_2^+$ fragments\cite{gaire2014,vacher2022,tilborg2009,ludwig2016,lucchini2022}. To avoid the signal from the former overlapping with that from the latter in the time-of-flight mass spectra, we used $^{13}$C-substituted ethylene, $^{13}$C$^{12}$CH$_4$, as the target, since this provided an unambiguous way for detecting the signal from the doubly-charged ions. The $^{13}$C$^{12}$CH$_4^{++}$ ions observed at a mass/charge ratio of $14.5$ were free from any contamination from other fragment ions of the parent molecule. 
The XUV-only and NIR-only time-of-flight (TOF) mass spectra for the molecule are shown in Fig.~\ref{fig2}A for the region-of-interest. In both cases no signal from the dication can be observed, demonstrating the simultaneous role of both pulses in producing the dication. We show the typical two-color signal of the dication as a function of the XUV pump - NIR probe delay and its mass-integrated counterpart in Fig.~\ref{fig2}B and C, respectively. 
The latter is fitted with a time-delayed Gaussian function (see Supporting Information for details about the data analysis) after precise determination of the time zero of the pump-probe delay with cross-correlation measurements in argon (see Supporting Information for details). Within the temporal resolution of the experiment, the yield does not show any exponential decay; rather it seems to be confined inside a narrow temporal window. 
The measurement has been repeated at several probe intensities and with different gases for the HHG process. As shown in Fig.~\ref{fig2}D, the delay $\Delta t$ at which the dication signal peaks, does not seem to exhibit any significant dependence on the NIR-probe intensity. The average value was found to be $\Delta t=14.8\pm1.7$ fs. Overcoming the binding-energy-gap between the singly-charged ion excited to any of the cationic states from D$_1$ to D$_4$ and the doubly-charged ion in its ground state S$_0$, requires between $9$ to $11$ NIR-photons. The time-of-flight mass spectra for probe-only ionization of the neutral molecule contained signal from the parent $^{13}$C$^{12}$CH$_4^{+}$ cation, at all NIR-probe intensities reported here.

\begin{figure*}[!htb]
\centering
\includegraphics[width=1.0\linewidth]{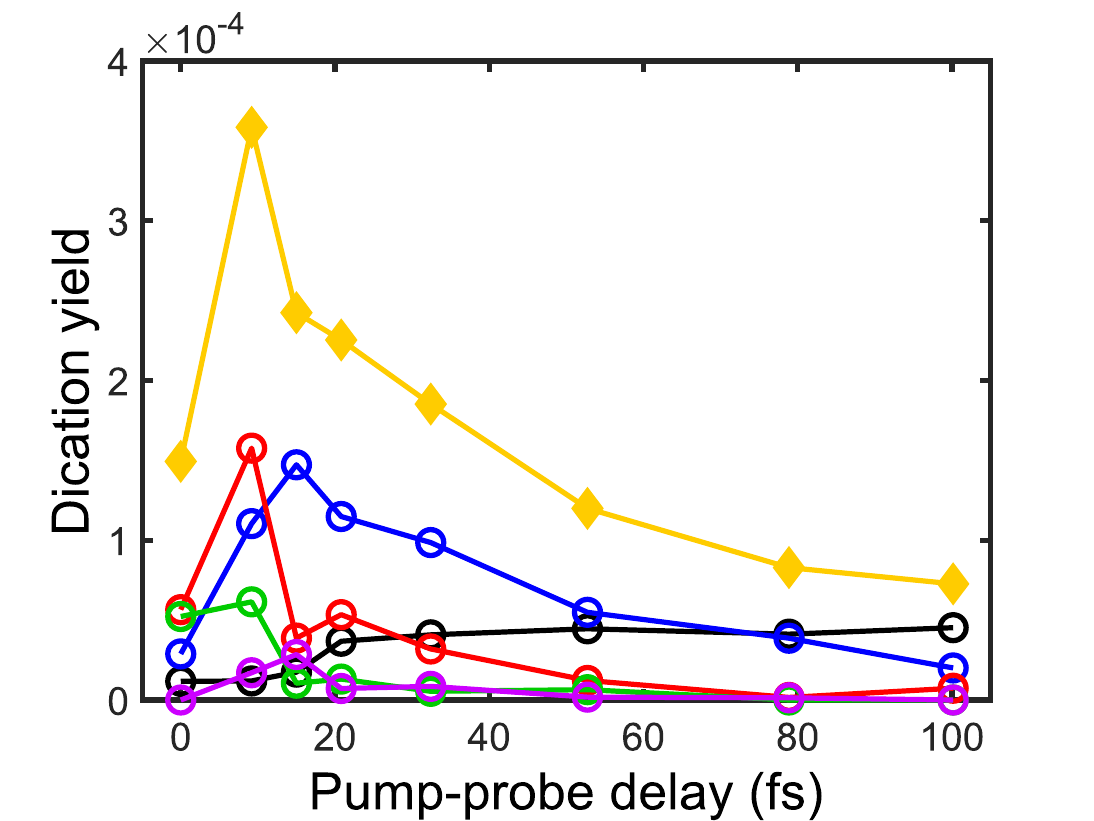}
\caption{Theoretical ethylene dication production probabilities, termed as yields (open circles) as a function of the pump-probe delay, when the cation is ionized from the D$_0$ (black), D$_1$ (blue), D$_2$ (red), D$_3$ (green), and D$_4$ (purple) electronic states. The yellow, solid diamonds denotes the combined contributions from all electronic states as a function of the delay. Note the excellent qualitative agreement with the experimental results reported in Fig.~\ref{fig2}C.}
\label{fig3}
\end{figure*}

Nevertheless, it is worth noting that our experiment operates firmly in the multi-photon regime as the intensity of the NIR-probe (between $10^{12}$ to $10^{13}$ W/cm$^{2}$) is more than one order of magnitude smaller than what is required for the tunneling ionization of molecular ions \cite{constant1996,wu2012,gong2014}.
The Keldysh parameter ($\gamma$) provides a qualitative measure for distinguishing between the multi-photon ($\gamma\gg 1$) and tunneling ionization ($\gamma\ll 1$) regime\cite{keldysh1965}. Given that the ionization energy gap between the ground state of the dication and the excited cationic states varies between $14.4$ and $17.45$ eV, the parameter $\gamma$ in the present case ranges from $3.5$ to $6$, indicating that our experiment was performed in the multi-photon regime.

To gain insight on how the XUV-induced dynamics in the cation and the subsequent effect of the multi-photon probe can lead to the appearance-delay in the dication yield in Fig.~\ref{fig2}C, we described the dynamics induced upon XUV-pump ionization with advanced trajectory surface-hopping simulations\cite{fdez-galvan2019,li-manni2023}, and explicitly took into account the effect of the multi-photon nature of the probe via a close-coupling based approach to molecular ionization\cite{randazzo2023}. The energy-gap between the different cationic states and the ground state of the dication entails the absorption of several NIR-photons, for which one-photon ionization probability calculations alone are insufficient. Indeed, multi-photon transitions are highly non-linear processes that are sensitive both to the presence of intermediate resonant states and to the ionization potential. While REMPI can increase the ionization rate, reduction in the ionization potential, and hence the number of photons necessary to ionize the system, also increases the ionization rate. Given these constraints, the description of the multi-photon ionization process even at a qualitative level requires explicit integration of the time dependent Schr\"odinger equation (TDSE) for the molecular system in the presence of the NIR-probe field. We achieved it by using the newly developed ASTRA package (AttoSecond TRAnsitions), which employs the formalism of transition-density matrices to accurately calculate photoionization observables for ions treated at the same CAS (Complete Active Space) electronic structure level as with the surface-hopping simulations\cite{randazzo2023}.

The average dication yield, $\langle\mathcal{Y}_i(t)\rangle$, from the $D_i$ cationic state were obtained from the following expression:
\begin{equation} 
\langle\mathcal{Y}_i(t)\rangle=\frac{\sum_{j}\mathcal{W}_{j}^{\text{XUV}}\mathcal{P}_{i}^{j}(t)\mathcal{Y}_{i}^j(t)}{\sum_{j}\mathcal{W}_{j}^{\text{XUV}}\mathcal{P}_{i}^{j}(t)}.
\label{equn1}
\end{equation}
Here, $\mathcal{W}_{j}^{\text{XUV}}$ is the XUV-induced ionization probability for the neutral ethylene molecule to the cationic state $D_j$, with $j\in\{0,1,2,3,4\}$, obtained by projecting the time-dependent wave function onto a complete set of single-ionization scattering states, defined by the aforementioned cations (see Supporting Information for details). The quantity $\mathcal{P}_{i}^{j}(t)$ denotes the fraction of the surface hopping trajectories initiated on the electronic state $D_j$ that are in electronic state $D_i$ at time $t$ fs. Subsequently, the time-dependent weight for the electronic state population can be expressed as $\langle w_i(t)\rangle=\sum_{j}\mathcal{W}_{j}^{{\text{XUV}}}\mathcal{P}_{i}^{j}(t)$. The quantity, $\mathcal{Y}_{i}^{j}(t)$, calculated using the ASTRA code, denotes the ionization yield from the $D_i$ ionic state, calculated at the average D$_{2h}$-symmetry geometry at pump-probe delay of $t$~fs, given that it was in $D_j$ ionic state at $0$~fs. Multiplying $\langle\mathcal{Y}_i(t)\rangle$ with $\langle w_i(t)\rangle$ results in the dication yield (probability of dication production) as a function of the pump-probe delay, as shown in Fig.~\ref{fig3} for $7$~TW~cm$^{-2}$ probe-intensity. The yield is explicitly dependent on the respective cationic population. The combined dication yield with contributions from all the relevant cationic states (solid diamonds) exhibits a peak at around $10$~fs, in excellent qualitative agreement with the experimental trend observed in Fig.~\ref{fig2}C. The dication yield, as reported in Fig.~\ref{fig3} increases by more than twice between $0$ and $10$~fs. It remains higher than that at $0$~fs until a pump-probe delay of $20$~fs, beyond which it decreases monotonically. Closer inspection reveals that the initial increase in the dication yield originates mainly from D$_2$ and D$_1$ cationic states, followed by minor contributions from D$_3$. A similar enhancement in the dication yield at around $10$~fs has also been observed at lower probe intensities, $3$~TW~cm$^{-2}$ and $5$~TW~cm$^{-2}$ (see Fig.~S1 in Supporting Information).

\begin{figure*}[!htb]
\centering
\includegraphics[width=1.0\linewidth]{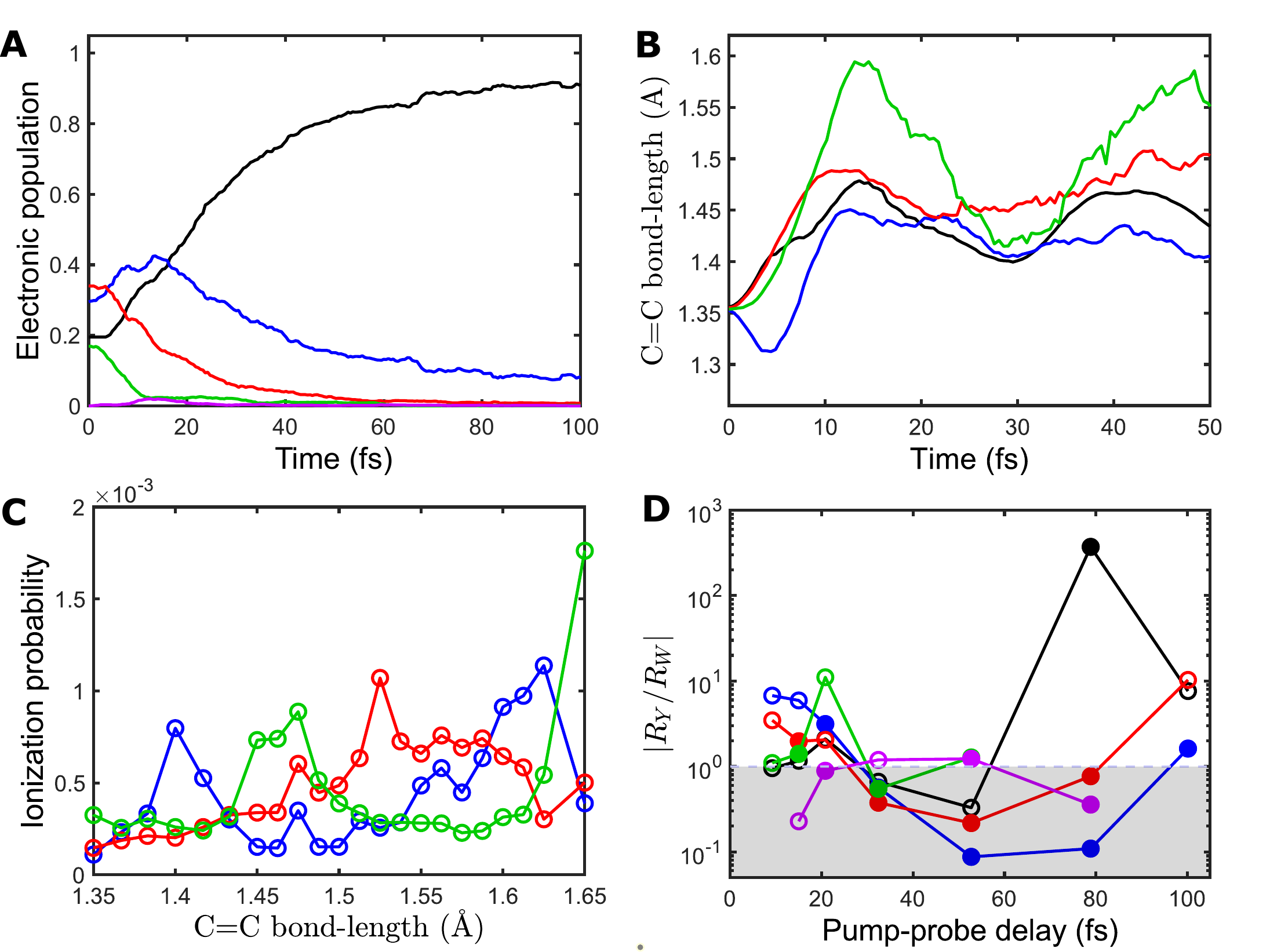}
\caption{({\bf A}) Temporal evolution of the electronic population for different cationic states (black: D$_0$; blue: D$_1$; red: D$_2$; green: D$_3$ and purple: D$_4$) weighted by the XUV-induced initial-state populations in D$_0$, D$_1$, D$_2$ and D$_3$. ({\bf B}) Temporal evolution of C=C bond lengths per current electronic state following initial ionization to D$_0$, D$_1$, D$_2$, and D$_3$. The results are weighted both by the XUV-ionization yields for each of the initial states, and by the populations of each current state at each point in time (see main text for details). ({\bf C)} Ionization probability (open circles) as a function of the C=C bond length from D$_1$ (blue), D$_2$ (red) and D$_3$ (green) electronic states. ({\bf D}) $|R_Y/R_W|$ as a function of pump-probe delays. The empty (full) circles indicate an increase (decrease) of the dication yield as a function of the delay. The time delay corresponds to the higher one in the interval analyzed. The time delays for which the yield vanishes either at the beginning or at the end of the time interval are excluded in to order avoid giving the false impression that both $\langle\mathcal{Y}_i(t)\rangle$ and $\langle w_i(t)\rangle$ contribute the same to the yield variation. In panels ({\bf C}) and ({\bf D}), the NIR-probe intensity was $7$~TW~cm$^{-2}$.}
\label{fig4}
\end{figure*}

To understand and disentangle the effects from the electronic and nuclear relaxation dynamics in the dication yield, first we looked closely at the time-dependent weighted average of the electronic populations, $\langle\mathcal{P}_{i}(t)\rangle$ of the cationic states $D_i$, obtained via the trajectory surface-hopping simulations in combination with multi-reference electronic structure calculations (see Supporting Information for details). Similar to Eq.~\ref{equn1}, the weighted average of the electronic populations is given by, 
\begin{equation} 
    \langle\mathcal{P}_i(t)\rangle=\frac{\sum_{j}\mathcal{W}_{j}^{\text{XUV}}\mathcal{P}_{i}^{j}(t)}{\sum_{j}\mathcal{W}_{j}^{\text{XUV}}}.
    \label{equn2}
\end{equation}
We considered contributions from the first four cationic states D$_0$, D$_1$, D$_2$ and D$_3$. This allowed us to study not only the effect of the ionization of the XUV-pump, but also the outcome of the subsequent ultrafast non-adiabatic dynamics. As can be seen in Fig.~\ref{fig4}A, within the first $10-20$ fs, the electronic population in the ethylene cation reaches a maximum for the lower-lying D$_1$ electronic state. One can expect higher-lying cationic states such as D$_3$ or, D$_4$ to contribute significantly to the dication yield due to their reduced energy-gap with the ground state S$_0$ of the dication (see Fig.~\ref{fig1}). However, this is not the case due to significant loss of excited state populations for the higher-lying cationic states.

Of the various internal nuclear coordinates, C=C bond stretching is the most likely to affect NIR ionization efficiency without also causing the fragmentation of the cation. Changes in other nuclear coordinates, such as the C-H bond length or, the H-C-H dihedral angle might lead to H-loss or, 2H/H$_2$-loss across comparable timescales\cite{vacher2022,tilborg2009,ludwig2016,lucchini2022}, altering the parent molecule. Relaxation dynamics in the ethylene cation mediated by the dissociation of the carbon-carbon double bond usually takes place over longer ($>50$~fs) timescales\cite{vacher2022}. Combining all these previous observations, we focused exclusively on the C=C bond stretching degree of freedom to explain the peak in the time-dependent dication yield from our combined surface-hopping and multi-photon ionization simulations.
Fig.~\ref{fig4}B shows the average C=C bond lengths, $\langle\mathcal{R}_i(t)\rangle$, for the cationic states from D$_0$ to D$_3$. These average values were obtained by replacing $\mathcal{Y}_{i}^{j}(t)$ in Eq.~\ref{equn1} with $\mathcal{R}_{\text{CC},i}^{j}(t)$, the mean distance between the two carbon atoms for trajectories that are in D$_i$ at $t$ fs, upon initial excitation to D$_j$. The carbon-carbon distance reaches a maximum between $10-20$ fs, with the largest change being in the case of D$_3$ electronic state, followed by that for the D$_2$ electronic state. 
The ionization probabilities from the most relevant cationic states (D$_1$ to D$_3$) in the doubly-charged ethylene production, as a function of the intercarbon distance, are shown in Fig.~\ref{fig4}C. These results were obtained by employing ASTRA, at a probe intensity of $7$~TW~cm$^{-2}$. Calculations for two other probe intensities, namely $3$ and $5$~TW~cm$^{-2}$ can be found in Fig.~S2 of Supporting Information. For all three states, the corresponding variation shows a peak: around $1.4$~\AA~for D$_1$, $1.52$~\AA~for D$_2$ and $1.47$~\AA~for D$_3$. The fact that the maximum appears at nearly the same C=C bond lengths within an approximate pump-probe delay of $10-20$~fs indicates an increase in the dication yield within the said time-window (see, Fig.~\ref{fig4}B). It is worth noting that the increase in ionization probabilities as a function of the C=C bond expansion is neither monotonic or, step-wise monotonic, which would be expected since the ionization potential decreases monotonically with the C=C distance surge and thus, it is possible to ionize the cation with less photons. Instead, they are sharply peaked at those particular C=C bond lengths, indicating the presence of REMPI (see Fig.~\ref{fig4}C). Although D$_3$ undergoes substantial C=C bond stretching and a pronounced decrease in the ionization potential, it contributes little to the observed peak in the dication yield due to its low residual population. 

To quantify the contributions from the electron and nuclear dynamics in the measured dication yield evolution, we have constructed two additional parameters in terms of $\langle w_i(t)\rangle$ and $\langle \mathcal{Y}_i(t)\rangle$ defined above. They are as follows: $R_w=\overline{\langle \mathcal{Y}\rangle}\Delta\langle w\rangle/Y$ and $R_{\mathcal{Y}}=\overline{\langle w\rangle}\Delta\langle \mathcal{Y}\rangle/Y$. Here, $\overline{\langle \mathcal{Y}\rangle}=[\langle \mathcal{Y}_i(t_f)\rangle+\langle \mathcal{Y}_i(t_0)\rangle]/2$ and $\overline{\langle w\rangle}=[\langle w_i(t_f)\rangle+\langle w_i(t_0)\rangle]/2$ are the average values for $\langle \mathcal{Y}_i\rangle$ and $\langle w_i\rangle$ calculated at two different pump-probe delays, with $t_0<t_f$. Similarly, $\Delta\langle \mathcal{Y}\rangle=[\langle \mathcal{Y}_i(t_f)\rangle-\langle \mathcal{Y}_i(t_0)\rangle]$ and $\Delta\langle w\rangle=[\langle w_i(t_f)\rangle-\langle w_i(t_0)\rangle]$ are the corresponding temporal differences in values for $\langle \mathcal{Y}_i(t)\rangle$ and $\langle w_i(t)\rangle$. The denominator, $Y=\overline{\langle w\rangle}\Delta\langle \mathcal{Y}\rangle+\overline{\langle \mathcal{Y}\rangle}\Delta\langle w\rangle$ is the total yield variation in the aforementioned interval, so that $R_w+R_{\mathcal{Y}}=1$. The absolute ratio, $|R_\mathcal{Y}/R_w|$ is shown in Fig.~\ref{fig4}D for NIR probe-intensity of $7$~TW~cm$^{-2}$. When $|R_\mathcal{Y}/R_w|>1$, the change in $\langle \mathcal{Y}_i(t)\rangle$ contributes more to the variation of the dication yield than the change in $\langle w_i(t)\rangle$, indicating a variation of the ionization rate itself due to nuclear relaxation. When $|R_\mathcal{Y}/R_w|<1$, the reverse is true, rather indicating a variation of the corresponding electronic state population due to electronic decay. 
The empty (full) circles indicate an increase (decrease) of the dication yield, $\langle w_i(t)\rangle\langle \mathcal{Y}_i(t)\rangle$. The results highlight that, indeed, the peak-like localization observed in the dication yield corresponds to strong contributions from ionization of the D$_1$ and D$_2$ cationic states to the dicationic ground state, S$_0$ via the multi-photon NIR-pulse. A similar trend has been observed for the lower probe intensities as well (see Fig.~S3 in Supporting Information). Fig.~\ref{fig4}D suggests an increase at short time delays due to an increase of the ionization rates because of nuclear relaxation, followed by a decrease due to electronic population decays. According to the trajectory surface hopping calculations, within a pump-probe delay of $10-20$~fs, only these two electronic states are appreciably populated, even though the XUV-ionization substantially populates several cationic states, from D$_0$ to D$_3$ (see, Fig.~\ref{fig4}A). 
The TDSE-based calculations taking into account the effect of the multi-photon probe indicate that the dication yield increases due to the C=C bond reaching a favorable length, resulting in a REMPI-type process, thus unveiling the state-selective nature of the NIR-probe. The exact symmetry of the electronic state responsible for the REMPI is difficult to determine, given that it depends on the potential energy landscape of the photo-excited cation as well as the exact intensity of the probe pulse at the instant of ionization.  Selectivity of the NIR probe was previously observed for XUV-induced dynamics in cations of polycyclic aromatic hydrocarbons, where the origin of selectivity was identified as the localization or delocalization of the vibrational modes involved in the relaxation dynamics spanning over several tens of femtoseconds\cite{boyer2021}. Compared to that, in the present case the electronic population for all excited cationic states decreases considerably beyond $30$~fs. While D$_0$ is gaining population after this delay, its ionization probability is so low compared with the other excited states that it only contributes to the yield's background, overall decreasing dication yields. 

We have shown that it is vital to include both the effect of the XUV-pump and the multi-photon NIR-probe when studying ultrafast dynamics in photo-excited molecular cations. On the one hand, the surface hopping simulations allow the description of the coupled electron and nuclear dynamics following XUV-ionization. On the other hand, the interpretation of the two-color signal obtained as a function of the pump-probe delay requires to include the ionization from the NIR-probe. The results presented here involves a model system, such as ethylene, allowing us to showcase the importance of combined tools to describe the observed femtosecond enhancement as well as confinement of the dication yield. While similar observations have previously been made in the tunneling ionization regime, our study highlights how ionization yield can be enhanced via a combination of relaxation of the electronic population as well as a change in the nuclear coordinate in photoionized species affecting the binding-energy gap. With the advent of intense XUV and x-ray sources providing ultrashort attosecond pulses, our study will motivate future experiments to study molecular multi-photon ionization processes at even shorter wavelengths.

\section{Author Information}
\subsection{Corresponding Authors}
*(L.A.) E-mail: luca.argenti@ucf.edu \\
*(M.V.) E-mail: morgane.vacher@univ-nantes.fr \\
*(S.N.) E-mail: saikat.nandi@univ-lyon1.fr
\subsection{Author Contributions}
C.M., L.F. and A.B. contributed equally to this project. A.B., V.L., F.L. and S.N. performed the experiment. C.M. performed the ab-initio calculations based on ASTRA package of codes under the supervision of L.A.. The trajectory surface hopping simulations were carried out by L.F. under the supervision of M.V.. C.M., L.F., L.A., M.V., and S.N. interpreted the results. S.N. wrote the manuscripts with inputs from L.A. and M.V. All authors discussed the manuscript. S.N. proposed and led the project.
\subsection{Notes}
The authors declare no competing financial interest.

\begin{acknowledgement}
We acknowledge the assistance of Emilien Prost during the experiment. Simulations in this work were performed using HPC resources from CCIPL (Le centre de calcul intensif des Pays de la Loire) and from GENCI-IDRIS (Grant 2021-101353). 
We acknowledge financial support from Agence National de la Recherche (ANR-20-CE29-0021 and ANR-21-CE30-0052). S.N. thanks Centre National de la Recherche Scientifique (CNRS) and F\'{e}d\'{e}ration de Recherche Andr\'{e} Marie Amp\`{e}re, Lyon for financial support. 
L.A. and C.M. acknowledge support of the DOE CAREER Grant No. DE-SC0020311. C.M. acknowledges the support by the National Science Foundation MPS-Ascend Postdoctoral Research Fellowship under Grant No. 2402225.
M.V. and L.F. received financial support under the EUR LUMOMAT project and the Investments for the Future program ANR-18-EURE-0012. L.F. acknowledges thesis funding from the Région Pays de la Loire and Nantes University. The project is also partly funded by the European Union through ERC Grant No. 101040356 (M. V.). The views and opinions expressed are however those of the authors only and do not necessarily reflect those of the European Union or the European Research Council Executive Agency. Neither the European Union nor the granting authority can be held responsible for them. 
\end{acknowledgement}

\begin{suppinfo}
The Supporting Information is available free of charge at
\begin{itemize}
  \item Description of the experimental set-up, data analysis, and the cross-correlation measurements using photoionization in argon atoms. Details about the non-adibatic dynamics simulations using trajectory surface hopping and sequential multi-photon ionization process using ASTRA code.
\end{itemize}
\end{suppinfo}

\end{document}